# Distinguishing the Rigidity Dependences of Acceleration and Transport in Solar Energetic Particles

**Donald V. Reames**

Institute for Physical Science and Technology, University of Maryland, College Park, MD 20742-2431 USA, email: dvreames@umd.edu

**Abstract** In solar energetic particle (SEP) events, the power-law dependence of element abundance enhancements on their mass-to-charge ratios *A/Q* provides a new tool that measures the combined rigidity dependences from both acceleration and transport. Distinguishing between these two processes can be challenging. However, the effects of acceleration dominate when SEP events are small or when the ions even propagate scatter-free, and transport can dominate the temporal evolution of large events with streaming-limited intensities. Magnetic reconnection in solar jets produces positive powers of *A/Q* from +2 to +7 and shock acceleration produces mostly negative powers from -2 to +1 in small and moderate SEP events where transport effects are minimal. This variation in the rigidity dependence of shock acceleration may reflect the non-planer structure, complexity, and time variation of coronal shocks themselves. Wave amplification by streaming protons in the largest SEP events suppresses the escape of ions with low *A/Q*, creating observed powers of *A/Q* from +1 to +3 upstream of the accelerating shock, decreasing to small negative powers downstream. Of course, the powers of *A/Q* are correlated with the spectral indices of He, O, and Fe, yet unexplained departures exist.

## 1. Introduction

The processes of acceleration and transport of solar energetic particles (SEPs) can depend upon particle velocity and magnetic rigidity. What is often called particle "energy," $E$, quoted as MeV amu$^{-1}$, is actually a measure of velocity $E = \mathcal{E}/A = M_u(\gamma - 1) \approx \frac{1}{2} M_u \beta^2$, where $\mathcal{E}$ is the total kinetic energy, $A$ is the atomic mass, $M_u = m_u c^2 = 931.494$ MeV, $\gamma = (1-\beta^2)^{-1/2}$, and $\beta = v/c$ is the particle velocity relative to the speed of light, $c$. The magnetic rigidity, or momentum per unit charge, is $P = pc/Qe = M_u \beta\gamma A/Q$ in units of MV, and determines an ions magnetic interactions and scattering by Alfvén waves, for example. When we measure the abundances of different elements at a constant velocity, this rigidity dependence causes relative enhancements in SEP ion abundances that often vary as a power-law function of their mass-to-charge ratio $A/Q$, relative to their average or source abundances in the corona. This power-law dependence was first noticed by Breneman and Stone (1985) for the elements with atomic numbers $6 \leq Z \leq 30$. These observations have been extended for many SEP events in recent years, and especially by including H and He (*e.g.* Reames, 2019b, 2020), but the relative contributions of acceleration and transport have not been considered previously. In many theories the rigidity dependence of either acceleration or transport is considered as an arbitrary adjustable parameter, offering little discrimination or physical insight. To what extent can we use the observations to distinguish the rigidity dependence of acceleration in an SEP event from that of transport away from the source?

What can the rigidity dependence of shock acceleration, for example, tell us? If shocks were simple planer structures, acceleration would be faster for ions with short mean free paths $\lambda$ as they can scatter back and forth across the shock more often, but shocks are now known to be complex surfaces, modulated by waves that can vary in space and time (*e.g.* Trotta *et al.*, 2020), and small-scale variations in interplanetary shocks have been observed by the *Cluster* spacecraft (*e.g.* Kajdič *et al.*, 2019). These variations also involve the angle between the magnetic field and the shock normal, $\theta_{Bn}$. Even the early review article by Jones and Ellison (1991) discusses the reduced rigidity dependence observed at the Earth's bow shock where shock smoothing could allow ions with longer mean free paths to preferentially encounter a larger velocity difference that compensates, as shown by Monte Carlo calculations (Ellison, Möbius, and Paschmann, 1990). What are the local rigidity dependences of the SEPs and what are the shapes of coronal shocks that drive their acceleration? How much do they vary? The purpose of this article is to begin a discussion of that subject for SEP observations.

The two primary types of SEP events exist that were originally designated "impulsive" or "gradual." They are now distinguished by an extensive history (*e.g.* Reames, 1988, 1995b, 1999, 2013, 2015, 2017; Gosling, 1993). In smaller "impulsive," or $^3$He-rich, SEP events (Mason, 2007; Bučík, 2020), acceleration is understood to occur at open magnetic-reconnection sites in solar jets (Kahler, Reames, and Sheeley, 2001; Bučík *et al.*, 2018a, 2018b; Bučík, 2020) from which the SEPs easily escape. Here, element abundances, relative to those in the corona, are observed to increase steeply as a power of $A/Q$, by a factor of ≈1000 across the periodic table from H or He all the way up to elements near Au and Pb (Reames, 2000; Mason *et al.*, 2004; Reames and Ng, 2004; Reames, Cliver, and Kahler, 2014a, 2014b) as a result of the reconnection process (*e.g.* Drake *et al.*, 2009). On average, abundance enhancements vary as $(A/Q)^{3.64 \pm 0.15}$ (Reames, Cliver, and Kahler, 2014a) with powers in individual events ranging from +2 to +7, with $A/Q$ determined at a temperature $T \approx 3$ MK in impulsive SEP events (Reames, Meyer, and von Rosenvinge, 1994; Reames, Cliver, and Kahler, 2014a, 2014b).

In contrast, the larger "gradual" SEP events (Lee, Mewaldt, and Giacalone, 2012; Desai and Giacalone, 2016), involve acceleration at shock waves, driven by fast, wide coronal mass ejections (CMEs) as shown by observations (Kahler *et al.*, 1984; Mason, Gloeckler, and Hovestadt, 1984; Reames, Barbier, and Ng, 1996; Cliver, Kahler, and Reames, 2004; Gopalswamy *et al.*, 2012; Cohen *et al.*, 2014; Kouloumvakos *et al.*, 2019), usually involving acceleration from ambient coronal material (Reames, 2016, 2020), supported by theories and models (Lee, 1983, 2005; Zank, Rice, and Wu, 2000; Ng and Reames, 2008; Afansiev, Battarbee, and Vainio, 2016; Hu *et al.*, 2017, 2018). Coronal abundances sampled by SEPs differ from photospheric abundances by a factor that depends upon the first ionization potential (FIP) of the element. High-FIP (>10 eV) elements travel across the chromosphere as neutral atoms while low-FIP elements are initially ionized and are preferentially enhanced a factor of ≈3 by the action of Alfvén waves (Laming, 2015; Reames, 2018a; Laming *et al.*, 2019); these FIP-processed coronal ions form the reference abundances sampled much later as SEPs (Webber, 1975; Meyer, 1985; Reames, 1995a, 2014, 2018a; Mewaldt *et al.*, 2002). It is interesting that the FIP abundance pattern of SEPs always differs from those of the solar wind (Mewaldt *et al.*, 2002; Desai *et al.*, 2003; Reames, 2018a, 2020; Laming *et al.*, 2019), probably involving different sources from open or closed loops when the atoms transit of the chromosphere to the corona. Shocks presumably continue to reaccelerate those same ions selected initially at 2 – 3 solar radii (*e.g.* Reames, 2009). SEPs are not accelerated solar wind.

After acceleration, ion scattering during transport can also depend upon a power of the ion magnetic rigidity (Parker, 1963), creating abundances with a power-law dependence upon *A/Q* for ions when compared at a constant velocity. In the largest gradual SEP events, this scattering is often induced by self-generated waves produced by the streaming protons themselves (Ng, Reames, and Tylka, 1999, 2001, 2003, 2012). During transport, for example, since Fe scatters less than O, Fe/O is enhanced early in an event and therefore depleted later, producing a power-law dependence upon *A/Q* that increases early and decreases later.

The distinction between impulsive and gradual SEP events becomes complicated when local shock waves at CMEs in jets are fast enough to reaccelerate the impulsive ions pre-accelerated in the magnetic reconnection. In addition, large pools of impulsive suprathermal ions can collect from many jets near active regions, producing $^3$He-rich, Fe-rich background levels that are often seen (Desai *et al.*, 2003; Bučík *et al.*, 2014, 2015; Chen *et al.*, 2015). These pools of impulsive suprathermal seed ions are preferentially accelerated by some shock waves (Desai *et al.*, 2003; Tylka *et al.*, 2005; Tylka and Lee, 2006) and can even dominate SEPs with weaker shocks or quasi-perpendicular shocks that may sometime favor or even require this boosted injection, while stronger shocks are more-easily seeded by ambient coronal plasma. Shocks probably always accelerate some impulsive suprathermal ions, when they are available (Mason, Mazur, and Dwyer, 1999), but for strong shocks they can only be enough for small fraction of the resulting SEPs.

Reames (2020) has identified four paths for producing observed SEP element abundances: (i) the "pure" impulsive events, SEP1, (ii) impulsive events with reacceleration by a local shock, SEP2, (iii) gradual events dominated by pre-accelerated impulsive seed ions, SEP3, and (iv) strong gradual events dominated by seed ions from the ambient corona plasma, SEP4.

SEP ion measurements used herein are from the *Wind* spacecraft near Earth. During most SEP events we can observe the relative abundances of the elements H, He, C, N, O, Ne, Mg, Si, S, Ar, Ca, and Fe as well as groups of heavier elements up to Pb from the *Low-Energy Matrix Telescope* (LEMT; von Rosenvinge *et al.*, 1995) on *Wind*. Abundances are taken primarily from the 3.2–5 MeV amu$^{-1}$ interval on LEMT, although H is only available near 2.5 MeV amu$^{-1}$. Abundance enhancements are measured relative to the average SEP abundances listed by Reames (2017; see also Reames, 1995a, 2014, 2020). We reference properties from the list of impulsive SEP events published in Reames, Cliver, and Kahler (2014a), and the list of gradual SEP events published in Reames (2016).

The process for determining the best-fit source plasma temperature and power-law in *A/Q* has now been described in many articles (*e.g.* Reames, Cliver, and Kahler, 2014b, 2015; Reames, 2016, 2019a, 2019b, 2019c, 2020), in review articles (Reames, 2018b), and a textbook (Reames, 2017). Basically, intensities of 14 ion species are accumulated in 8-hour intervals and are divided by corresponding reference abundances, mentioned above, to produce enhancements. Temperatures *T* in nine intervals from 0.8 to 6 MK are considered. Each temperature determines well-defined theoretical *Q* values for each ion species (*e.g.* Mazzotta *et al.*, 1998; Post *et al.*, 1977) allowing a power-law fit of enhancement *versus A/Q* and determines $\chi^2/m$ for the fit at that *T*. Minimum values in plots of $\chi^2/m$ *versus T*, which are shown in many of the figures, select the best-fit *T* and the best power of *A/Q* for that 8-hour period. These are plotted throughout the event; usually these source plasma temperatures remain fairly constant during an event even as abundance patterns change. The behavior of H and He is more complex, so fits are done for elements with $6 \leq Z \leq 56$ and are extrapolated to protons at *A/Q* = 1. For most gradual SEP events the protons fit this extrapolated line (these are SEP4 events), but for gradual SEP events with impulsive seed particles (SEP3 events) there is a large proton excess above the fit line which has been taken to mean an additional contribution from protons accelerated from the ambient solar plasma (Reames, 2019b, 2019c, 2020). Temperatures inferred from abundances have been a boon to SEP studies since direct measurements of SEP ionization states are extremely rare.

## 2. Scatter-Free Events

Those SEP events that permit the easiest determination of the rigidity dependence of acceleration are the small events whose ions propagate to us with negligible scattering. The term "scatter free" is generally applied when the scattering mean free path is $\geq 1$ AU. These scatter-free events are primarily the small $^3$He-rich events where this transport was first quantified and studied in some detail by Mason *et al.* (1989).

Figure 1 shows the slope or power of *A/Q* for impulsive events of different size as measured by peak intensities of 2.2 MeV protons and indicates the speed of any associated CMEs. The dashed line in Figure 1 is at the typical intensity of the $^3$He-rich events studied by Mason *et al.* (1989) corrected for difference in energy. Even moderately large impulsive SEP events show anisotropic angular distributions (*e.g.* Reames, Ng, and Berdichevsky, 2001).

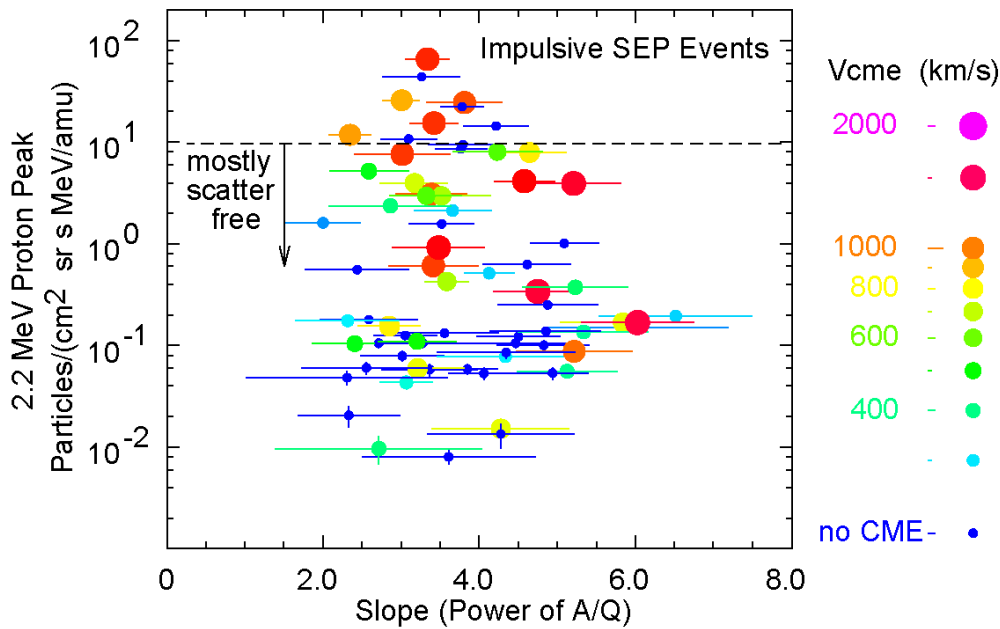

**Figure 1** Peak 2.2-MeV proton intensities are shown *versus* the slope (*i.e.* power of *A/Q*) for impulsive SEP events with measurable proton intensities. The associated CME speed is indicated for those events with observed CMEs (data from Reames, 2019a)

Those events in Figure 1 with high intensities and CME speeds above about 500 km s$^{-1}$ are typically classed as SEP2 events where local shock waves have reaccelerated ions originally accelerated from the magnetic reconnection. The smaller events and especially many of those without visible CMEs are more likely to be the "pure" SEP1 events from the reconnection site.

Figure 1 tentatively indicates an upper bound for the intensity of scatter-free events. Of course, scattering does depend upon the pre-existing condition of the interplanetary medium, but the average small scatter-free events studied by Mason *et al.* (1989) were below this limit. Thus most of the impulsive SEP events are probably scatter free and the acceleration by magnetic reconnection yields power-law slopes between about +2 and +7, a rather wide range, but are not significantly altered by transport. There is no evidence that the slopes of SEP2 events are systematically altered by additional acceleration by the local CME-driven shocks, except that the larger events seem to show slopes from +2 to +5, somewhat less variation in slope.

The situation is rather different for gradual events since there must be enough wave intensity and scattering at the shock for acceleration to occur. Thus, scatter-free transport from gradual events is rare. Nevertheless, small gradual SEP events, even those that accelerate ambient coronal plasma, are not impossible, as shown by the sample in Figure 2. The angular distributions of He and H shown in 2(c) and 2(d), respectively show minimal scattering of the ions during the first day of the event when abundances are collected to measure the power-law element enhancements *versus A/Q* shown 2(f). Fits were obtained with *A/Q* values defined by many different temperatures with the best $\chi^2/m$ value obtained at 1.0 MK as shown in 2 (d). The least-squares fit line is obtained for the elements with $6 \leq Z \leq 26$ and is extended down to protons. This analysis procedure has been described in more detail by Reames (2016, 2017, 2018b). It is more difficult to establish that a gradual event is scatter free since the source function is not impulsive but has an ill-defined extent in time, but the proton intensity is limited and angular distributions show minimal scattering. Thus, the slope of -1.07±0.14 certainly seems to be a measure of rigidity dependence of the acceleration at the shock in this event.

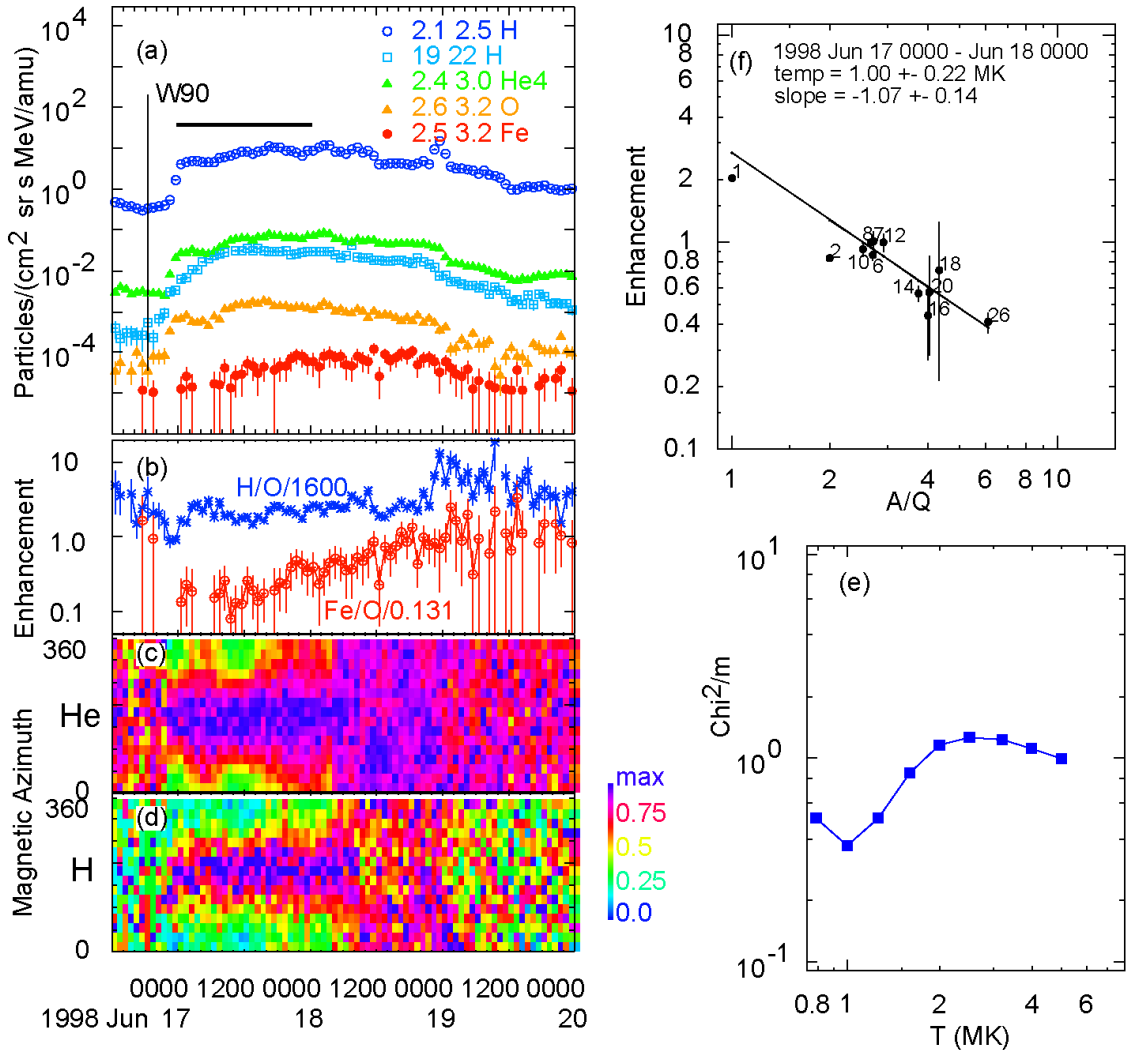

**Figure 2** Time variations of (**a**) intensities of the particles and energies listed in MeV amu$^{-1}$, (**b**) enhancements of H/O and Fe/O, and angular distributions (in degrees) relative to the magnetic field for (**c**) He and (**d**) H, are shown for the gradual SEP event of 17 June 1998. A power-law fit to the abundance enhancements for elements denoted by *Z* is shown in (**f**) for *A/Q* values at the best-fit temperature found at minimum $\chi^2/m$ shown in (**e**).

## 3. Modest Gradual Events

Power-law element enhancements *versus A/Q* for larger gradual SEP events are generally studied in 8-hr intervals since the powers can vary with time, especially during the largest events but generally not for small events, as we shall see. However, event-averaged slopes or powers of *A/Q* are used to show the overall pattern of gradual SEP event distribution in Figure 3.

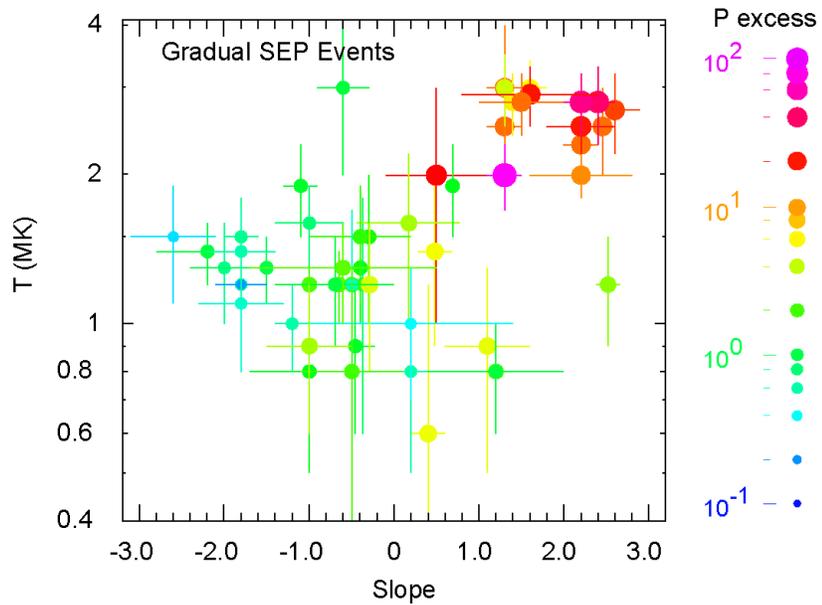

**Figure 3** The source plasma temperature is shown *versus* the average slope of the *A/Q* dependence for gradual SEP events with the average proton excess factor shown as size and color of the point (data from Reames 2019b).

The steep positive slopes and high proton excess identifies the SEP3 events where pools of suprathermal ions from impulsive events form the seed population that dominates the shock-accelerated SEPs along with excess protons sampled from the ambient plasma (Reames 2019b, 2020). For Figure 4 we remove these high-temperature events and show the fluence of >30 MeV protons *versus* slope for the remaining SEP4 events.

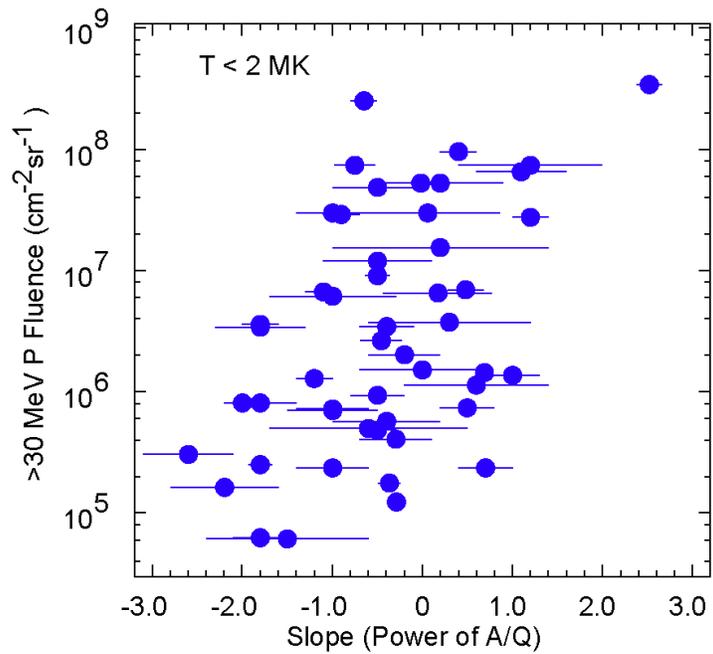

**Figure 4** The fluence of $\geq 30$ MeV protons is shown *versus* the average slope or power of the *A/Q* dependence for gradual events from $T < 2$ MK ambient plasma (SEP4 events).

While these events are all larger than the 17 June 1998 event in Figure 2, Figure 4 gives us a basis for selecting and studying the size range of gradual SEP events from the list in Reames (2016). We show analysis of two small gradual events that are rather different. Figure 5 shows analysis of the event of 4 June 1999 with a fairly low fluence of >30 MeV protons of $1.8 \times 10^5$ cm$^{-2}$ sr$^{-1}$. Figure 6 shows the event of 12 September 2000 with a fluence of $8.2 \times 10^5$ cm$^{-2}$ sr$^{-1}$.

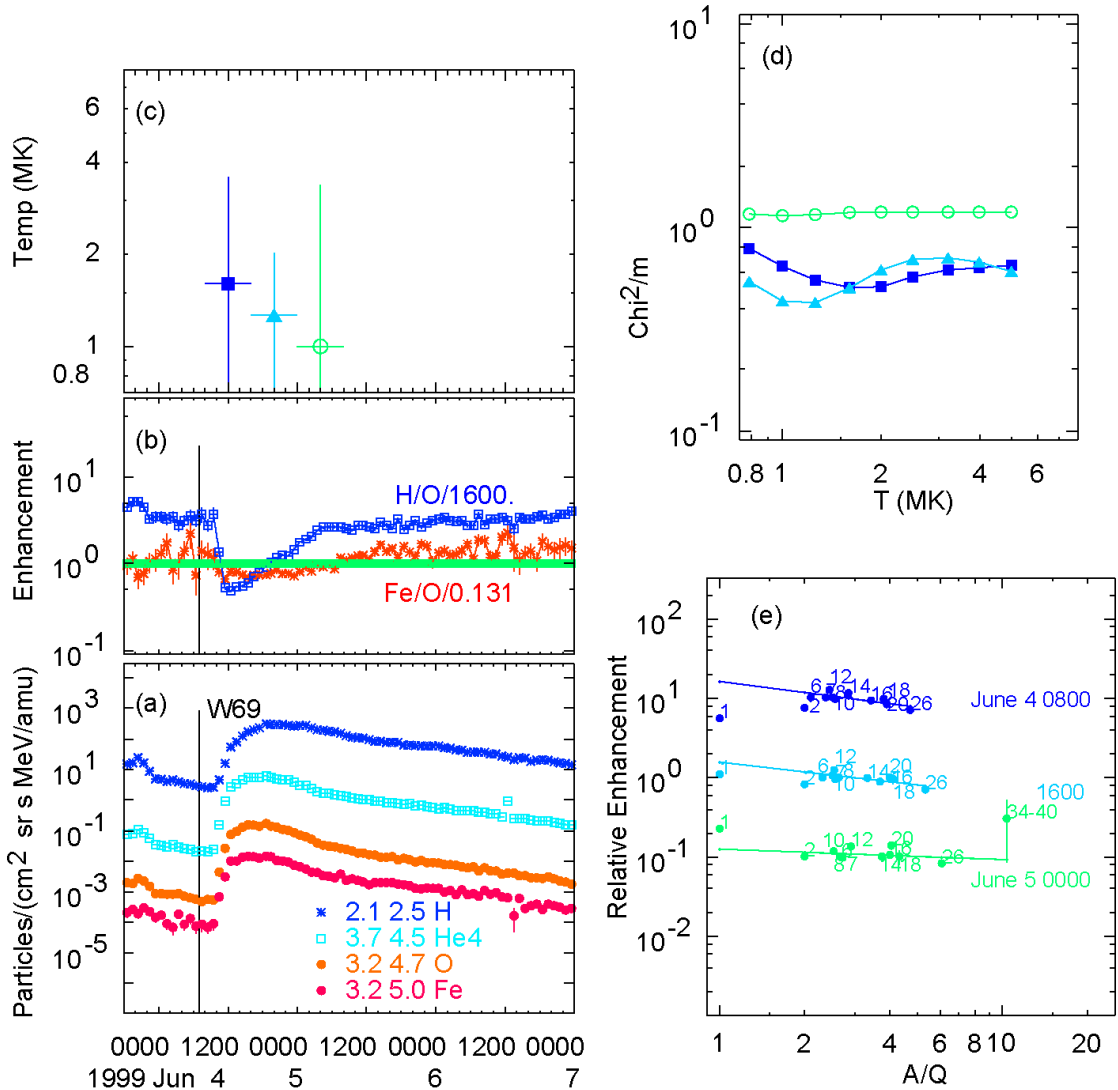

**Figure 5** (**a**) Intensities of listed ions and energies in MeV amu$^{-1}$, (**b**) enhancements, and (**c**) derived source temperatures are shown *versus* time for the 4 June 1999 gradual SEP event. (**d**) Shows $\chi^2/m$ *versus* $T$ where best fit $T$ provides $A/Q$ values for (**e**) fits of enhancements of elements, listed by $Z$, *versus* $A/Q$. Colors correspond for curves in (**c**), (**d**), and (**e**). Selection of $T$ is poor because dependence on $A/Q$ is flat.

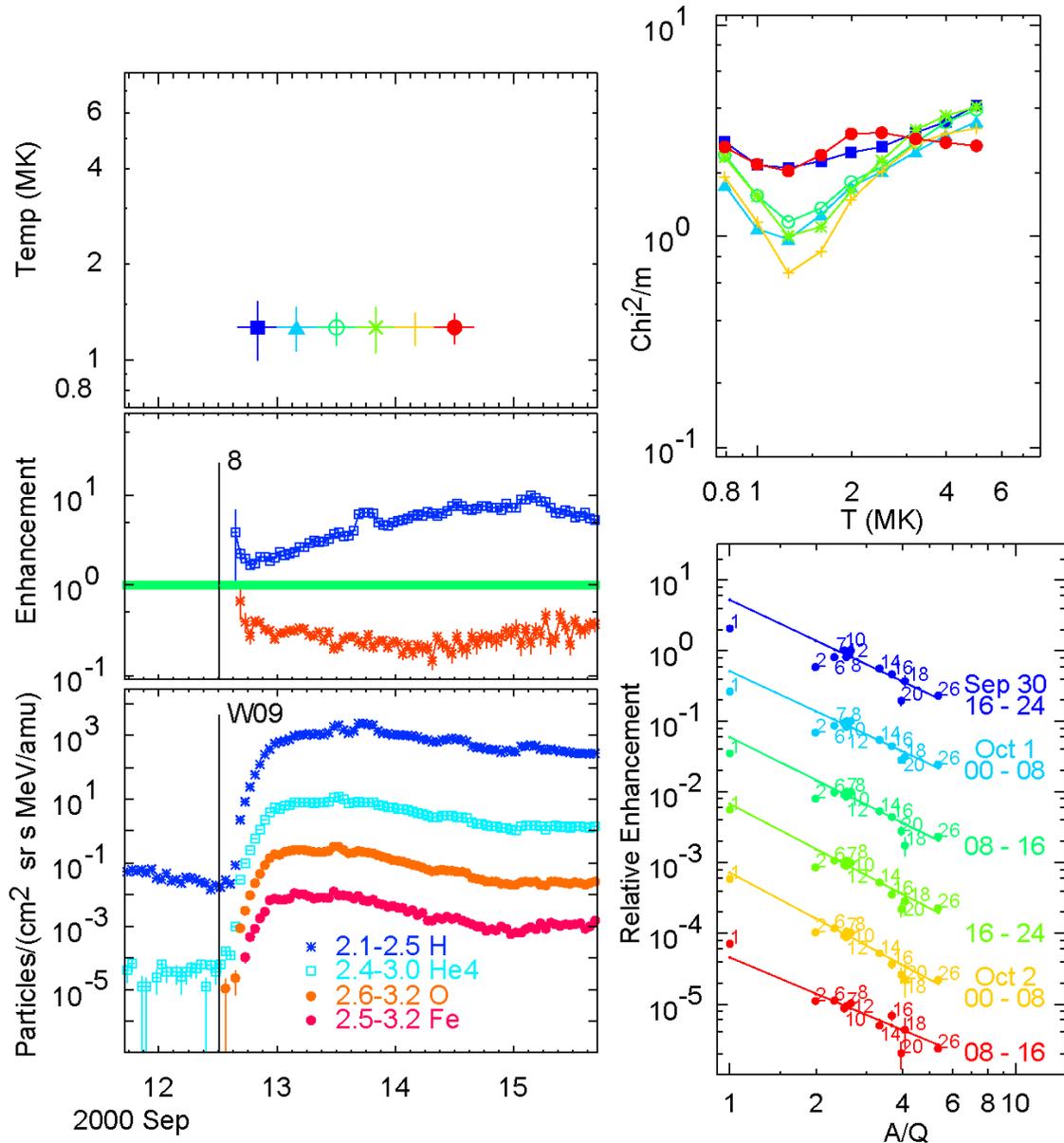

**Figure 6** (**a**) Intensities of listed ions and energies in MeV amu$^{-1}$, (**b**) enhancements, and (**c**) derived source temperatures are shown *versus* time for the 12 September 2000 gradual SEP event. (**d**) shows $\chi^2/m$ *versus T* where best fit *T* provides *A/Q* values for (**e**) fits of enhancements of elements, listed by *Z*, *versus A/Q*. Colors correspond for curves in (**c**), (**d**), and (**e**).

Despite its somewhat larger fluence of >30 MeV protons, the 12 September 2000 event in Figure 6 shows little change of the slope of *A/Q* with time from its average value of -2.0 ± 0.15. The modulation by scattering during transport would be expected to change with time and no such change is seen. Thus the dependence on *A/Q* probably all comes during acceleration. The smaller event of 4 June 1999 in Figure 5 has a slope of - 0.37 ± 0.12 that is so flat that $T$ is poorly defined, but this value must also come from the shock. If we examine the values in the lower decade of Figure 4, we might conclude that shocks can generate powers of $A/Q$ from about -2.5 to +1.0.

## 4. Large Gradual Events and the Streaming Limit

There is considerable evidence that intensities early in largest gradual events are limited by waves amplified by the high intensities of streaming protons, *i.e.* transport dominates (Reames 1990; Reames and Ng, 1998, 2010; Ng, Reames, and Tylka, 2012). This limit is also implicit in the steady-state streaming models of Bell (1978a, 1978b) and of Lee (1983, 2005). Figure 7 shows properties of the "Bastille-day" event of 14 July 2000 ($V_{CME}$ = 1674 km s$^{-1}$). The >30 MeV proton fluence for this event is $3.4 \times 10^8$ cm$^{-2}$ sr$^{-1}$. Figure 7(a) shows proton intensities up to 500 MeV and plateau energy spectra in Figure 7(f) where the transport of low-energy ions have been limited by scattering on waves generated by high intensities of protons above ≈10 MeV as they stream away from the shock (Reames and Ng, 2010; Ng, Reames, and Tylka, 2012).

Figure 7(e) shows power-law enhancements *versus A/Q* with positive slopes of +1.7 to +2.7 at times preceding the passage of the shock wave that drives the event and negative slopes of -0.6 to -0.25 after the shock passage. Clearly the ions with higher *A/Q* find it easier to penetrate the wave turbulence during transport out to 1 AU, increasing the slope of the observed power law. It is more difficult to say how much the negative power laws seen after the shock represent SEPs that have been depleted of high-*A/Q* ions.

Figure 8 shows the analysis of two more events for which streaming-limit plateau spectra have been studied, 4 November 2001 ($V_{CME}$ = 1810 km s$^{-1}$) on the left and 28 October 2003 ($V_{CME}$ = 2459 km s$^{-1}$) on the right. The top two panels show the power law fits in *A/Q* for each event. For the 4 November 2001 event the slopes are 0.8 to 2.0 before the shock and -0.8 to -0.6 after. For the 28 October 2003 event the slopes are 1.3 to 2.1 before the shock and -0.8 to -0.5 after.

The scattering that limits the intensities begins near the shock where the streaming intensities are highest, so it is difficult to separate acceleration from transport. Power-law dependence downstream of the shock is probably affected by depletion of the heavier ions.

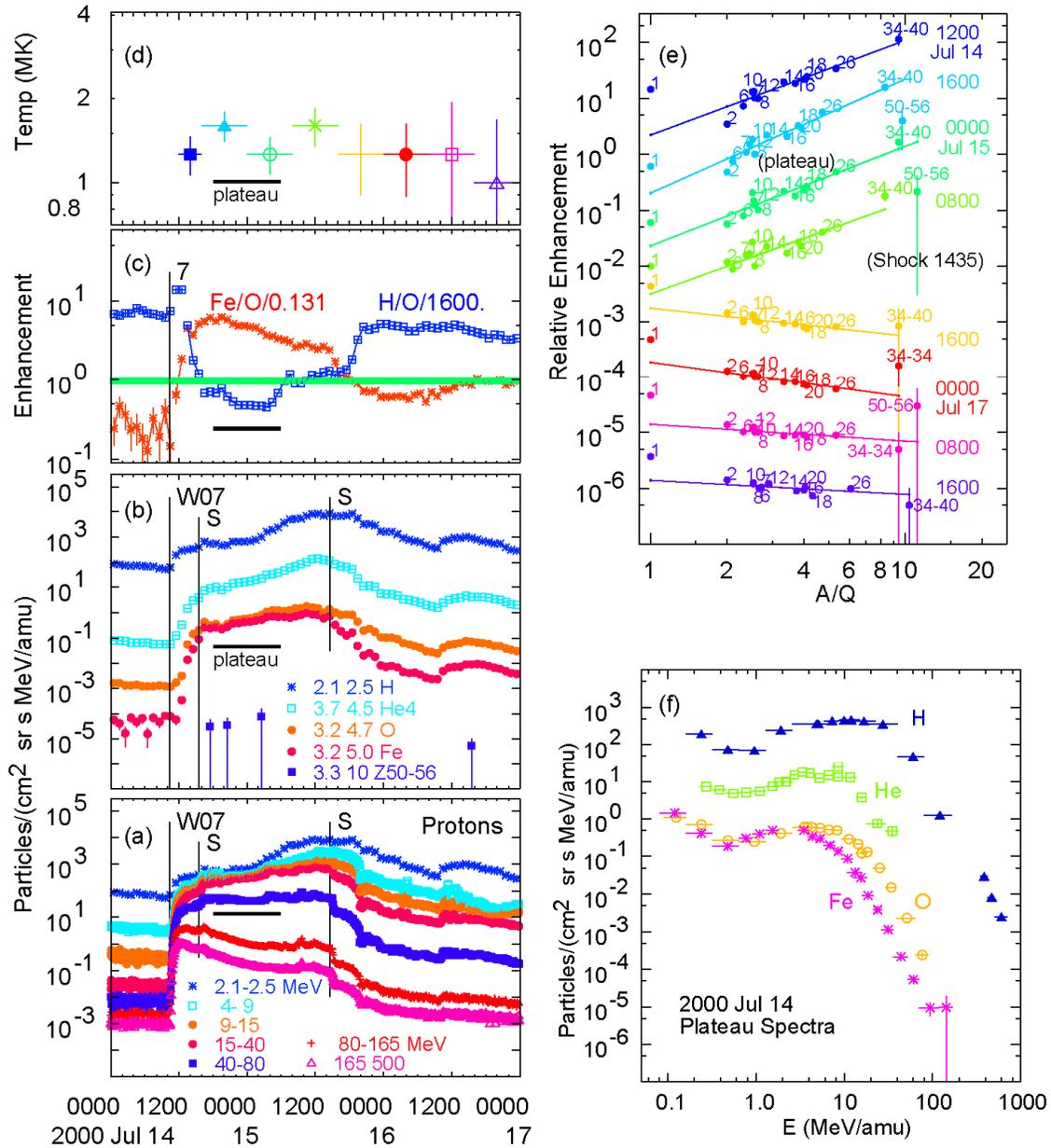

**Figure 7** (**a**) Proton intensities from 2 to 500 MeV, (**b**) selected ion intensities at energies in MeV amu$^{-1}$, (**c**) enhancements of H/O and Fe/O, and (**d**) derived source-plasma temperatures are shown *versus* time for the large 14 July 2000 gradual SEP event. In (**e**) least-squares fits of enhancements of elements, listed by *Z*, are shown *versus A/Q* for 8-hr intervals, colors corresponding with those in (**d**). Streaming-limited energy spectra in the marked plateau region are shown in (**f**).

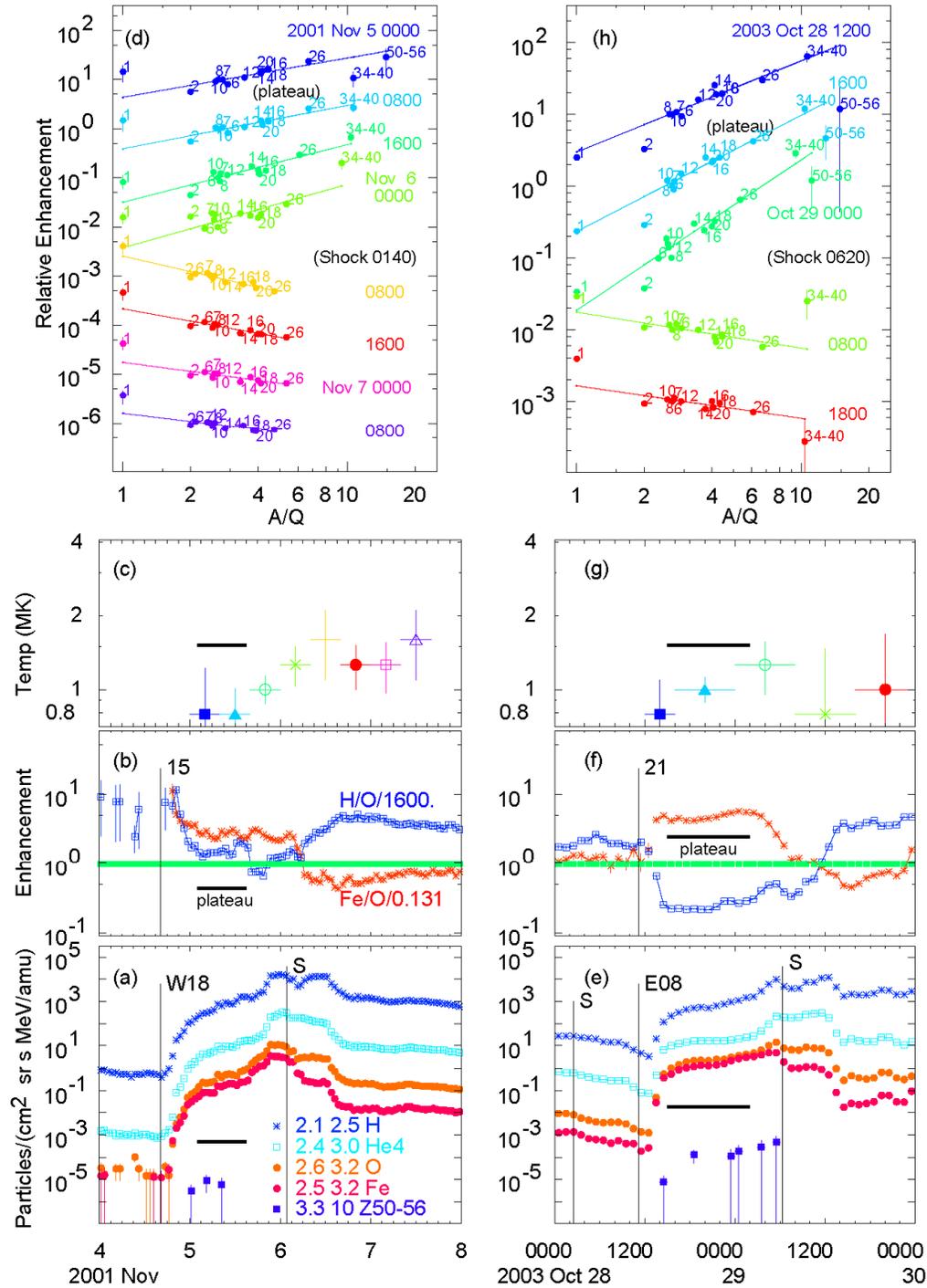

**Figure 8** Properties of two large gradual SEP events on (*left*) 4 November 2001 and (*right*) 28 October 2003 include relevant intensities of listed ions and energies in MeV amu$^{-1}$ in (**a**) and (**e**), enhancements of H/O and Fe/O in (**b**) and (**f**), and derived plasma temperatures in (**c**) and (**g**). Least-squares fits of enhancements of elements, listed by *Z*, are shown *versus A/Q* in (**d**) and (**h**), for 8-hr intervals with colors corresponding to those in (**c**) and (**g**), respectively.

## 5. Impulsive-Seeded Shock (SEP3) Events

In our study of gradual SEP events, we bypassed those gradual events where the SEPs with $Z \geq 2$ are dominated by ions with $T > 2$ MK believed to be reaccelerated from pools of impulsive suprathermal ions, the SEP3 events (Reames, 2020). The distribution >30 MeV proton fluence in these events is shown in Figure 9. Like other gradual events shown in Figure 4, many of these events are fairly small and have the distinctive positive power that is a property of their seed particles and does not change during the events.

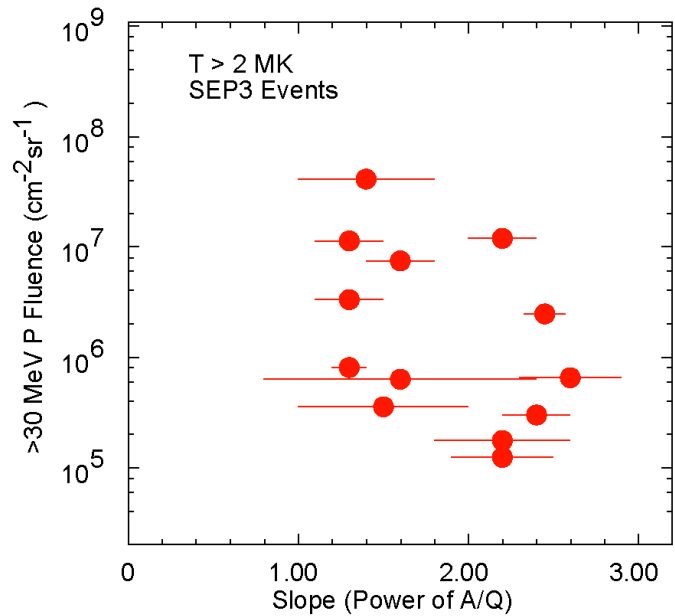

**Figure 9** The distribution of >30 MeV proton fluence *versus* the average slope or power of the *A/Q* dependence is shown for gradual SEP3 events dominated by ions from reaccelerated $T > 2$ MK impulsive suprathermal seed particles.

Figure 10 shows the analysis of an SEP3 event with a modest >30 MeV proton fluence of $8.7 \times 10^5$ cm$^{-2}$ sr$^{-1}$. The best-fit temperature shows a fairly constant value of 3.2 MK throughout the event in Figure 10(d) and the power-law fit for the $Z \geq 6$ enhancements has a positive slope that only varies from $+1.46 \pm 0.15$ to $+1.10 \pm 0.23$ during the event. Nevertheless, this is flatter that the average impulsive SEP event that contributes the seed particles, as is the case for all of the SEP3 events (Figure 9). If we assume that both H and He are derived from ambient plasma, the slope defined by the enhancements of these two elements is flat early then declines somewhat later; for other SEP3 events (*e.g.* Figures 5, 6, and 7 in Reames, 2019b) the slope from H to He is negative throughout.

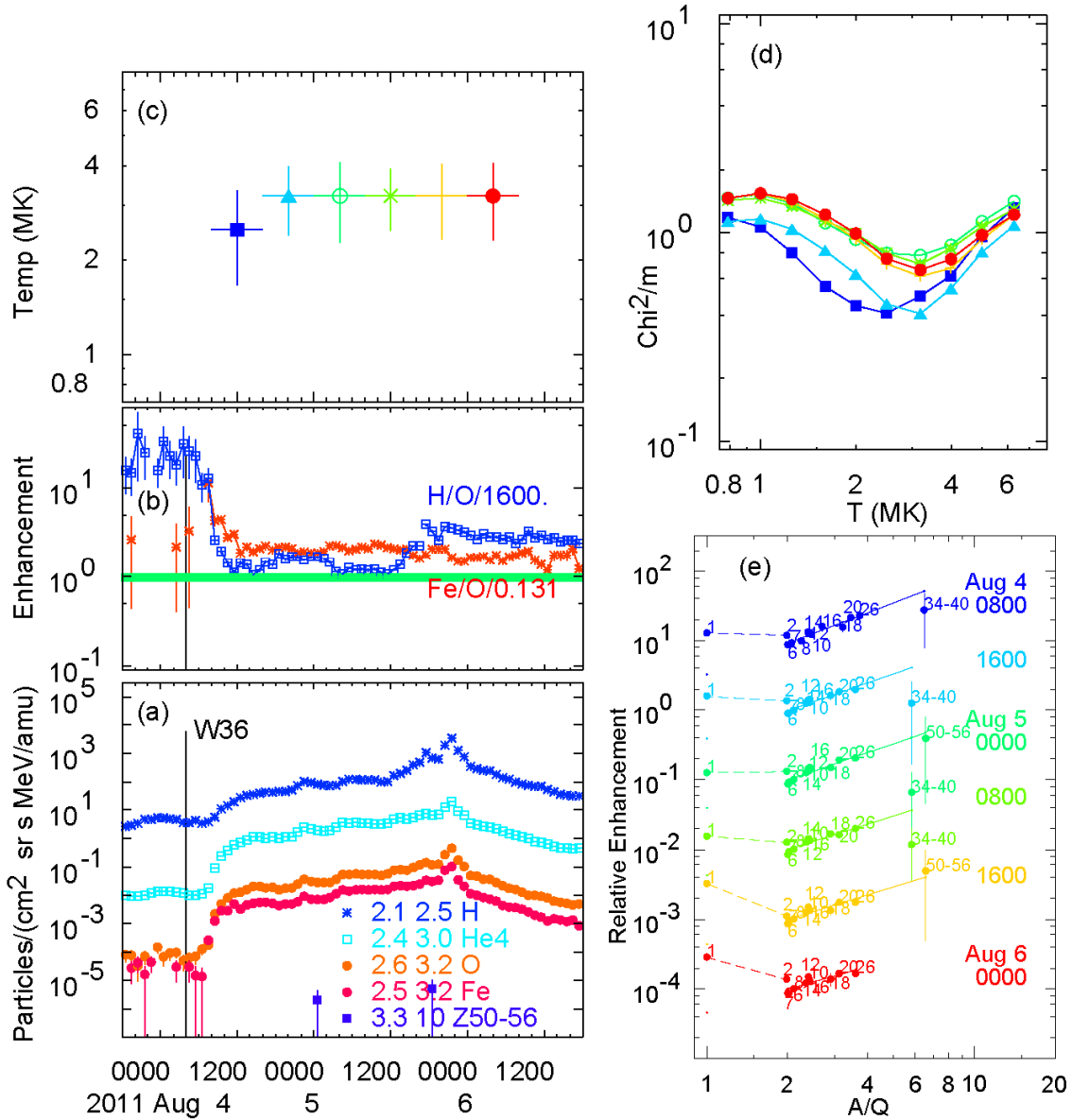

**Figure 10** (**a**) Intensities of listed ions and energies in MeV amu$^{-1}$, (**b**) enhancements of H/O and Fe/O, and (**c**) derived source temperatures are shown *versus* time for the 4 August 2011 gradual SEP3 event. (**d**) Shows $\chi^2/m$ *versus* $T$ where best fit $T$ provides $A/Q$ values for (**e**) fits of enhancements of elements, listed by Z, *versus* $A/Q$. Colors correspond for curves in (**c**), (**d**), and (**e**).

# 6. Energy Spectral Indices and Powers of *A/Q*

For shock acceleration, it is not surprising that the power of *A/Q* in an SEP event is related to the energy spectral indices of the ions. Figure 11 shows that the energy spectral index for O is highly correlated with the constant-velocity power of *A/Q* for 8-hr intervals during gradual SEP4 events that directly sample ambient coronal abundances, *i.e.* lack evidence of any impulsive suprathermal ions. The large events of Section 4 are included in Figure 11.

**Figure 11** The energy spectral index for oxygen is shown *versus* the power law of *A/Q* for the first four 8-hr intervals in SEP4-class gradual SEP events with $T < 2$ MK. Points are identified by event numbers (from list in Reames, 2016) shown in contrasting colors. Parameters of a linear least-squares (solid line) and expected (dashed line) fits are given.

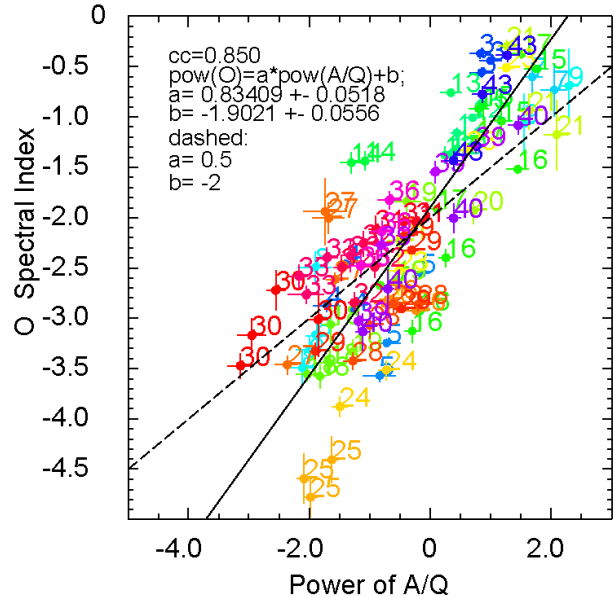

Examining the fit parameters in Figure 11, since $E$ varies as $p^2$, we would actually expect the parameter $a = 0.5$, and we might also expect that $b = -2.0$. However, local processes distort the expected relationship and some events even lie far off the fitted line; Events 3, 14, and 27 (from the list of Reames, 2016) lay far above the line and Event 25 far below. Thus, the correlation in Figure 11 permits a wide variation in power of *A/Q*. In some cases these differences correspond with differences among the spectral indices of He, O, and Fe (for comparisons of spectral indices, please see Reames, 2014). Is this just a problem with non-power-law spectra? If so, why? Do the differences depend upon shock structure or upon features of the Alfvén-wave spectrum?

Many small- and moderate-sized SEP4 gradual events assume the expected relationship of spectral indices and powers of *A/Q* as shown for the 24 August 1998 event in Figure 12. The figure shows details of the fitting and comparison of spectral indices of O and Fe with powers of *A/Q* for the event. The event has similar spectral indices for O and Fe (Figures 12(d), 12(e), and 12(f)), except for some degradation of the highest-energy Fe at the latest time (Figure 12(e)).

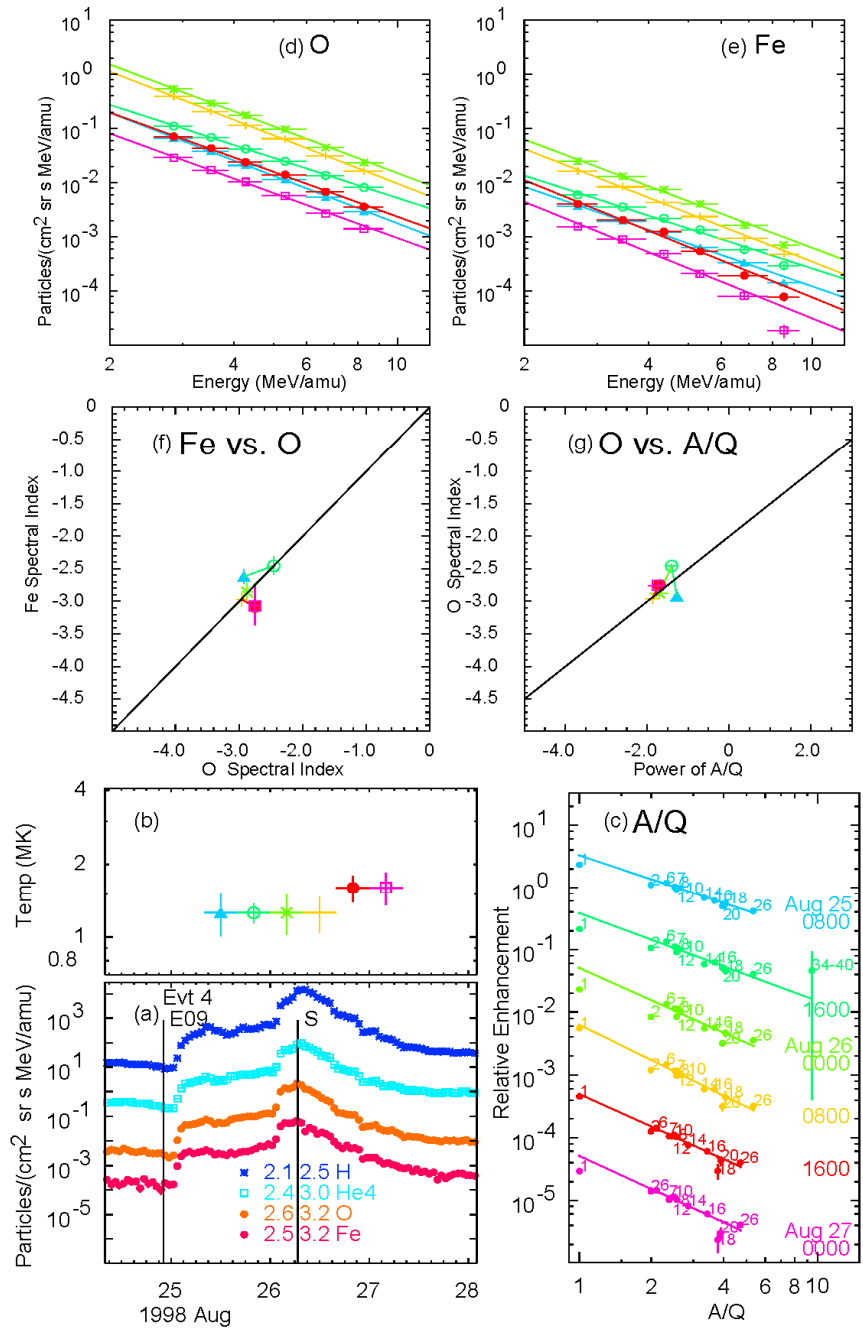

**Figure 12** (**a**) Intensities of listed ions and energies in MeV amu$^{-1}$, and (**b**) derived source temperatures are shown *versus* time for Event 4, the 24 August, 1998 gradual SEP event. Fits are shown for (**c**) enhancements of elements, listed by *Z*, *versus A/Q*, and for energy spectra of (**d**) O and (**e**) Fe. Correlation plots are shown for spectral indices of (**f**) Fe *versus* O and (**g**) of O *versus A/Q*. Colors for time intervals correspond in (**b**), (**c**), (**d**), (**e**), (**f**), and (**g**).

In contrast, Figure 13 shows comparison of spectral indices of O and Fe with the powers of *A/Q* for Event 14 (from the list of Reames, 2016) identified above as an outlier.

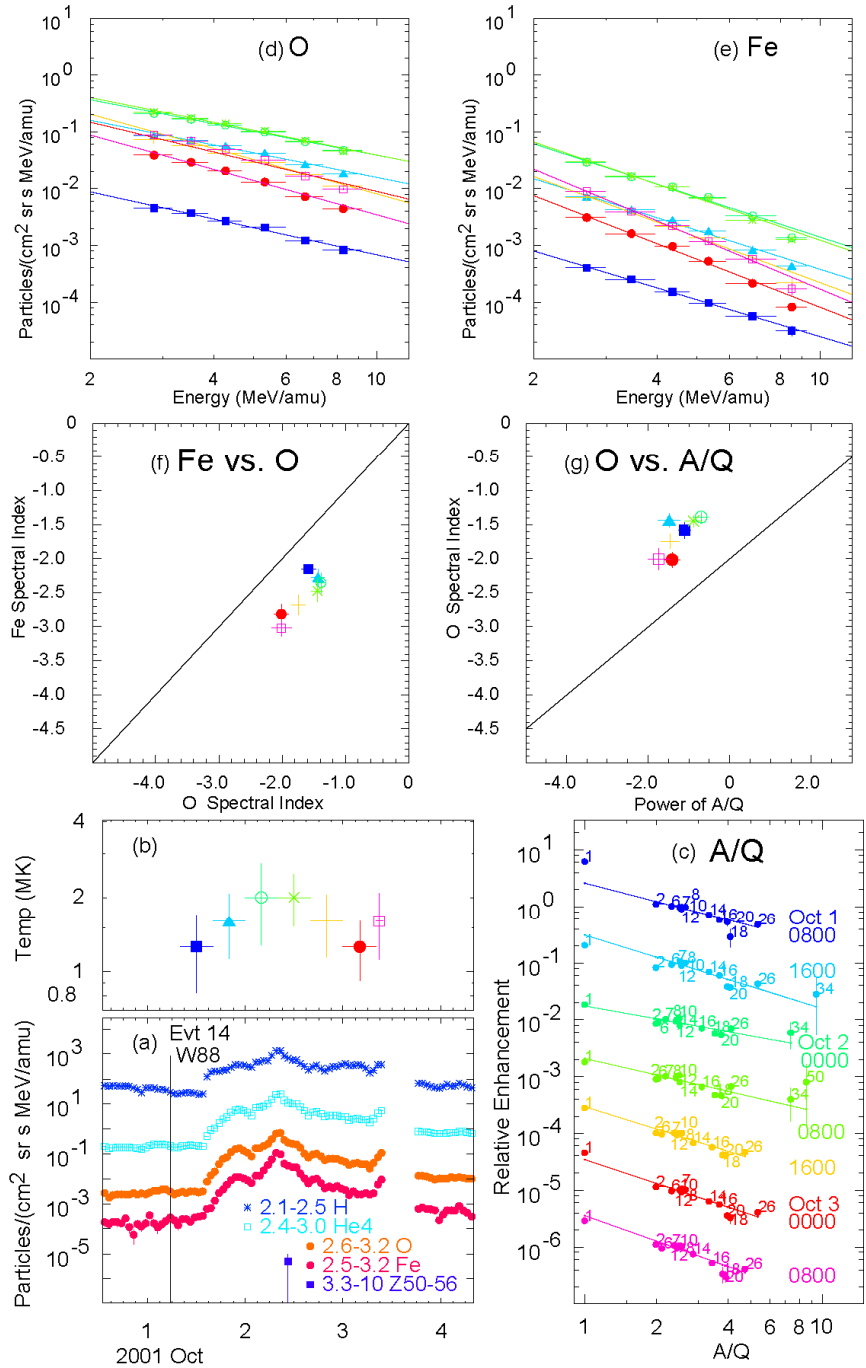

**Figure 13** (**a**) Intensities of listed ions and energies in MeV amu$^{-1}$, and (**b**) derived source temperatures are shown *versus* time for Event 14, the 1 October, 2001 gradual SEP event. Fits are shown for (**c**) enhancements of elements, listed by *Z*, *versus A/Q*, and for energy spectra of (**d**) O and (**e**) Fe. Correlation plots are shown for spectral indices of (**f**) Fe *versus* O and (**g**) of O *versus A/Q*. Colors for time intervals correspond in (**b**), (**c**), (**d**), (**e**), (**f**), and (**g**).

The available measurements of Fe and O in Figure 13 cover the same interval of energy nucleon$^{-1}$ (*i.e.* velocity) so that the Fe has a much higher rigidity. Panels 13(d), 13(e), and 13(f) show that the Fe spectra are steeper than the spectra of O for this event. Spectral indices for He (not shown) agree with those for O. Thus, differences in rigidity of the various measurements, and changes in the power law fits *versus* rigidity might explain variations for this event. This can often be especially true for very large events, for example in Figure 7(f), where the flattened spectra break downward above the plateau at a lower energy for Fe than for O, and O breaks downward at a lower energy than He, since these breaks tend to occur at a defined rigidity.

The lack of large time variations in the spectral indices and in the powers of *A/Q*, suggests minimal transport dependence for the events shown in Figures 12 and 13. Larger events with flatter spectra, such as event 3 noted above, have been analyzed in terms of wave generation during transport (Tylka, Reames, and Ng, 1999; Ng, Reames, and Tylka 1999). Most small- and moderate-sized SEP4 gradual events have reasonably well-correlated spectra of He, O, and Fe that agree with the expected values of the powers of *A/Q*. For SEP3 events, the power of *A/Q* is largely determined by the seed population, tends to remain constant, and is uncorrelated with the spectral indices that are determined by the shock. The wide variety of such relationships for differing SEP events will be explored further in a future publication.

## 7. Discussion and Conclusions

We have found that for most small and moderate SEP events the rigidity dependence or power of *A/Q* is almost entirely determined by the acceleration, with the transport contributing almost nothing at all. The impulsive SEP events have powers of *A/Q* ranging from +2 to +7 and perhaps from +2 to +5 for the larger events with associated CMEs fast enough to provide shock reacceleration.

Small and moderate gradual SEP events are also controlled by acceleration with very little time variation in powers of *A/Q* occurring during the events. In events dominated by residual impulsive seed particles with $T > 2$ MK, powers from *A/Q* from +1 to +2.5 are seen, suppressed by 1 or 2 powers from the average value of $\approx 3.6$ for the impulsive events themselves. Modest gradual events dominated by ambient coronal material with $T < 2$ MK have powers of *A/Q* ranging from -2.5 to +1. The cause of this large range in powers for similar small gradual SEP events is a considerable puzzle.

Shock acceleration with such a wide range of rigidity dependence may suggest substantial variation in shock structure, which may include shock smoothing (Jones and Ellison 1991) or much more turbulent variations that have been observed recently in interplanetary shocks near 1 AU on *Cluster* (*e.g.* Kajdič *et al.* 2019). The rigidity dependence of *λ* depends upon the spectrum of resonant waves each ion encounters, while the rigidity dependence of the acceleration may also depend upon the relative scale size of variations in the shock structure. At present we lack the theory or modeling that would allow us to associate specific shock structures with rigidity dependence and powers of *A/Q* in ions that probe it.

In large gradual SEP events, transport becomes increasingly important, but since the most intense wave growth and scattering begins just upstream of the shock, the processes of acceleration and transport may be coupled. In the largest SEP events powers of *A/Q* are 1 to 3 upstream of the shock and -0.8 to -0.2 downstream. The seed population for shock acceleration in these events is ambient coronal plasma. In the largest events, energy in SEPs can become comparable with that in the shock itself; these become SEP-mediated shocks which are even seen at 1 AU (*e.g.* Lario *et al.*, 2015; Mostafavi, Zank, and Webb, 2017). Models of shock acceleration and transport of SEPs with self-consistent wave generation have not yet been extended beyond protons.

Measurements of element abundance enhancements *versus A/Q* provide a new window on the structure of shocks contributing to acceleration of SEPs. Modeling the physics of these processes and extending this tool could increase the depth of our understanding of SEP acceleration and transport. Can models generate self-consistent SEP abundances and spectra?

## Disclosure of Potential Conflicts of Interest

The author declares he has no conflicts of interest.